# Spin glass behavior in an interacting $\gamma$-Fe$_2$O$_3$ nanoparticle system

D. Parker[*1], V. Dupuis[+2], F. Ladieu[1], J.-P. Bouchaud[1], E. Dubois[2], R. Perzynski[2], E. Vincent[1]

[1]Service de Physique de l'Etat Condensé (CNRS URA 2464)
DSM/DRECAM/SPEC – CEA Saclay
91191 Gif sur Yvette Cedex – France

[2]Laboratoire des Liquides Ioniques et Interfaces Chargées (LI2C)
UMR 7612 CNRS, Université Pierre et Marie Curie, ESPCI
Boîte 51, 4, place Jussieu 75252 Cedex 05 - France

**Abstract:** In this paper we investigate the superspin glass behavior of a concentrated assembly of interacting maghemite nanoparticles and compare it to that of canonical atomic spin glass systems. ac versus temperature and frequency measurements show evidence of a superspin glass transition taking place at low temperature. In order to fully characterize the superspin glass phase, the aging behavior of both the thermo-remanent magnetization (TRM) and ac susceptibility has been investigated. It is shown that the scaling laws obeyed by superspin glasses and atomic spin glasses are essentially the same, after subtraction of a superparamagnetic contribution from the superspin glass response functions. Finally, we discuss a possible origin of this superparamagnetic contribution in terms of dilute spin glass models.

* present address : Inorganic Chemistry Laboratory, Department of Chemistry, University of Oxford, South Parks Road, Oxford OX1 3QR, United Kingdom.
+ corresponding author: vdupuis@ccr.jussieu.fr

## 1. Introduction

Magnetic nanoparticle assemblies have attracted much attention over the last decade as promising media for high density magnetic recording [1]. For such applications, sufficiently large nanoparticles must be used to avoid spurious thermal relaxations of the magnetic moments, on which one wants to record the bits of information (a problem known as the "superparamagnetic limit" [2] in the field of magnetic recording), and to avoid complex effects due to disordered surface spins [3, 4]. Furthermore, concentrated assemblies of individually responding magnetic entities are required. This last requirement is difficult to satisfy as interparticle dipole-dipole interactions are strongly enhanced by both an increase in the size of the magnetic nanoparticles and an increase in their concentration [5]. At the present stage, a better understanding of the collective behavior of concentrated assemblies of magnetic nanoparticles is thus needed in order to address this problem.

Many studies in the past decades ([6-8] and references therein, for a review see [9]) have shown that increasing the nanoparticle concentration in magnetic nanoparticle assemblies yields a transition from a superparamagnetic state to a disordered collective state [10]. This state was called a superspin glass state by analogy with the disordered and frustrated magnetic state observed at low temperatures in spin glass materials [11]. Characteristic features of spin glasses such as a strong enhancement of magnetic non linearities [12-14] as well as dynamic scaling behavior [15-17] with reasonable values of the critical exponents have been observed close to the superspin glass transition temperature for a variety of nanoparticle systems. Dynamical studies [7] have revealed the existence of slow dynamics and aging in the superspin glass phase while more sophisticated protocols [18-24] have been used to illustrate the history dependent nature of these slow dynamics [25].

Despite the observed qualitative similarities, superspin glasses differ from canonical atomic spin glasses in several aspects. Firstly, the interacting magnetic moments have very different amplitudes ($10^2 - 10^4$ $\mu_B$ for strongly coupled spins in a single-domain magnetic nanoparticle compared to a few $\mu_B$ for an atomic spin) and the nature and range of their interactions are different (anisotropic and long range dipole-dipole interactions for magnetic nanoparticles vs. shorter ranged exchange or longer range RKKY interactions for atomic spins). Additionally the characteristic time of the spin flip mechanism is very short (~$10^{-12}$s) and nearly temperature independent in atomic spin glasses while it is much longer in superspin glasses, thermally activated and thus exponentially dependent on the ratio of the magnetic anisotropy energy $E_a$ to the thermal energy $k_BT$.

In this paper, we investigate the scaling behavior of the response functions of a superspin glass. Our main goal is to assess to what extent the scaling behavior observed in atomic spin glasses can be extended qualitatively and quantitatively to strongly interacting magnetic nanoparticle assemblies. The sample studied here is a frozen concentrated ferrofluid made of $\gamma$-$Fe_2O_3$ (maghemite) nanoparticles dispersed in water. We report observations which strongly suggest a transition towards a superspin glass state at low temperature as can be expected in such systems. We then investigate and quantitatively compare the aging and scaling behavior of the response functions in the superspin glass state to that of atomic spin glasses.

The paper is organized as follows. In section 2 we give details of the preparation and the basic characterization of the sample, in particular the experimental procedures used to probe the low temperature superspin glass transition and the scaling behavior of the response functions in the superspin glass phase. To allow comparisons with atomic spin glasses we also give a short introduction to the scaling laws obeyed by the response functions of atomic spin glasses. Section 3 is devoted to a presentation of our experimental results, first on the superspin glass transition and then on the scaling behavior of the response functions in the aging regime. Finally, in section 4 we propose new scaling laws that allow a consistent description of both dc and ac response functions and discuss their connection with theoretical results originally obtained in the context of dilute atomic spin glasses.

## 2. Experimental

### 2 (a) Sample preparation and characterization

The samples used for this study are well-defined and well-controlled magnetic colloidal dispersions of $\gamma$-$Fe_2O_3$ nanoparticles in water. In concentrated dispersions of this type, the interparticle magnetic interactions are high enough to obtain a glassy state at low temperatures while controlling the dispersion of the nanoparticles as well as the global interactions. The $\gamma$-$Fe_2O_3$ nanoparticles are chemically synthesized in water as described in reference [26]. Their surface is coated with citrate molecules that ensure a negative superficial charge at pH 7. Consequently, the particles can be dispersed in water at pH 7 due to an electrostatic interparticle repulsion that counterbalances the attractive interactions between particles [27]. Each particle is a nanometric monocrystal and a magnetic monodomain because the size of the particles is small enough to prevent the formation of Bloch walls. Thus each particle bears a permanent magnetic moment $\mu = m_s V$, where $m_s$ is the particle magnetization ($3.1 \times 10^5$ A/m) and $V$ is the particle volume [28] giving a typical particle magnetic moment of approximately $1.1 \times 10^4$ $\mu_B$. At low volume fraction of magnetic nanoparticles, the dispersions are superparamagnetic at room temperature with a magnetization curve following a Langevin formalism. The strong dependence of the magnetization on the size of the particles allows the characteristics of the size distribution to be determined. This distribution is well described by a log normal law, characterized by a mean diameter $d_0$ ($\ln d_0 = \langle \ln d \rangle$) and a polydispersity index $\sigma_d$, obtained from a two parameter fit of the experimental curves at room temperature measured at low volume fraction ($\Phi \sim 10^{-2}$ %).[29] The sample used here is characterized by $d_0 = 8.6$ nm and $\sigma_d = 0.25$.

In order to obtain superspin glasses, the dispersions must be concentrated. This is achieved using osmotic compression with a defined salt concentration [30]. At the end of the process, the volume fraction $\Phi$ of magnetic nanoparticles is measured. The sample used here is characterized by $\Phi = 35\%$ and a concentration of free sodium citrate in the dispersions of 0.03 mol/L. This sample is macroscopically solid at 300 K (it does not flow). As the sizes of the nanoparticles are polydisperse, this solid is not a crystal but a glass. This has been verified by Small Angle Neutron Scattering (SANS), which shows that the colloid has an amorphous structure. Given the experimental conditions, in particular the salt concentration, the interparticle potential is globally repulsive at 300 K, which gives a homogeneous dispersion of nanospheres without aggregates, as also confirmed by SANS [31].

These samples are chemically stable in time, however they dehydrate very quickly in air and therefore have to be protected accordingly. The material is cut into a parallelepiped (2 mm x 2 mm x 6 mm) of mass 59 mg and placed in a plexiglass cell, which is then sealed. The sample is kept inside this cell for the whole experiment, hence preventing water loss.

## 2 (b) Experimental methods

All the measurements reported here were carried out using a commercial Cryogenics S600 SQUID magnetometer in the temperature range 5 – 250K. In order to characterize the superspin glass transition we first measured the temperature dependence of the Field Cooled (FC) and Zero Field Cooled (ZFC) susceptibilities as well as that of the ac susceptibility (in phase and out-of-phase components) in the frequency range 0.04 Hz – 4 Hz (ac field amplitude 0.5 Oe). As strong magnetic non linearities are expected close to a spin glass transition we also measured the temperature dependence of the dc FC and ZFC susceptibilities for increasing probing fields in the range 0.3 – 60 Oe. Finally, to investigate the out-of-equilibrium dynamics in the superspin glass phase and test its scaling behavior, we performed thermo-remanent magnetization (TRM) and zero field cooled magnetization (ZFCM) as well as ac susceptibility relaxation experiments using experimental procedures widely used in spin glass studies [32] shown and described in Figure 1 (dc probing field of 0.5 Oe).

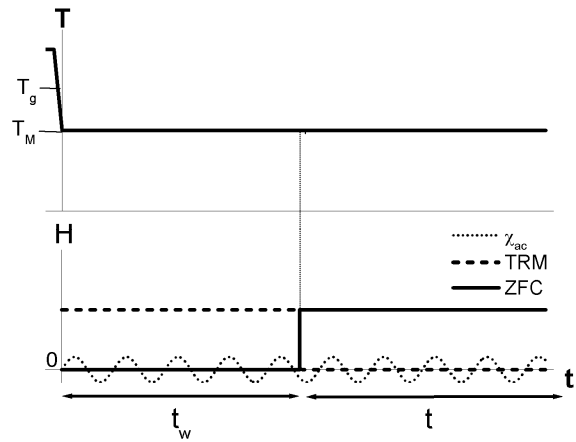

**Figure 1:** TRM (dashed line), ZFCM (full line) and AC relaxation (dotted line) measurement protocols. In the case of TRM and ZFCM measurements, the sample is heated to a temperature above the spin glass transition temperature ($T_g$) and subsequently cooled to the measuring temperature ($T_m$) in the presence of a small excitation field, $H$ (TRM) or in zero field (ZFCM). After waiting for a time, $t_w$, the field is respectively cut or applied and the relaxation of the magnetization is measured over a time, $t$. In the case of ac susceptibility relaxations, the sample is heated to a temperature above $T_g$ and subsequently quenched to the measuring temperature $T_m$. A weak ac magnetic field is applied throughout this procedure and the ac susceptibility is recorded as a function of the time elapsed since the quench.

## 2 (c) Scaling behavior of response functions of atomic spin glasses in the aging regime

The scaling behavior of response functions in atomic spin glasses has been widely investigated [32]. Here we only recall some of the basic results obtained in atomic spin glasses which will be used in the comparison of our superspin glass sample to atomic spin glasses.

In atomic spin glasses, the thermo-remanent magnetization following a quench in the spin glass phase can be written on general grounds and after a normalization to the field cooled magnetization as a sum of a stationary equilibrium part $m_{eq}(t)$ and an aging part $m_{ag}(t,t_w)$,

$$\frac{M}{M_{FC}} = m_{eq}(t) + m_{ag}(t,t_w) = A_{TRM}\left(\frac{\tau_0}{t}\right)^\alpha + f\left(\frac{t}{t_w^\mu}\right) \quad (1)$$

where $t$ is the time elapsed since the cutting of the field, $t_w$ is the waiting time (see Fig. 1), $\tau_0$ is an microscopic attempt time ("spin flip time"), $f$ is a scaling function, $A_{TRM}$ is a prefactor and $\alpha$ and $\mu$ are scaling exponents. It is noteworthy that the aging part obeys an approximate $t/t_w$ scaling ($\mu$ is usually ~ 1) as predicted in general theories of spin glasses [32]. More rigorously, the scaling variable which yields the best collapse of the $M/M_{fc}$ – $m_{eq}$ curves takes the more complicated form $\lambda/t_w^\mu$ where $\lambda$ is an effective time = $t_w^{1-\mu}$ $[(1+t/t_w)^{1-\mu} -1]/[1-\mu]$ which accounts for the evolution of the aging dynamics during the relaxation ($\lambda \sim t$ for $t<<t_w$, see details in reference [32]).

According to the above, and following linear response theory, the in-phase and out-of-phase components of the ac susceptibility after a quench in the spin glass phase can also be written as a sum of a stationary equilibrium part, $\chi_{eq}(\omega)$ and an aging part, $\chi_{ag}(\omega,t_w)$. As for the TRM, the aging part should follow an approximate $\omega t_w$ scaling and it is found that $\chi_{ag}(\omega, t_w) = A(\omega t_w)^{-b}$ where $b \sim 0.2$. A is an amplitude parameter which is found to be different for the in-phase and out-of-phase components of the ac susceptibility. Focusing on the reduced equilibrium parts, the following behavior should be expected from (1):

$$\frac{\chi'_{eq}(\omega)}{\chi'(\omega=0)} = \frac{\chi'(\omega,t_w) - \chi'_{aging}(\omega t_w)}{\chi'(\omega=0)} = 1 - A_{TRM}G(1-\alpha)Cos\left(\frac{\pi\alpha}{2}\right)\times(\omega\tau_0)^\alpha \qquad (2)$$

$$\frac{\chi''_{eq}(\omega)}{\chi'(\omega=0)} = \frac{\chi''(\omega,t_w) - \chi''_{aging}(\omega t_w)}{\chi'(\omega=0)} = A_{TRM}G(1-\alpha)Sin\left(\frac{\pi\alpha}{2}\right)\times(\omega\tau_0)^\alpha \qquad (3)$$

where G is the gamma function, $\omega$ is the ac field frequency and $\alpha$ is the scaling exponent already introduced in Equation 1.

In all the previous expressions, typical values of $\tau_0$ are $\sim 10^{-12}$ s (temperature independent) for atomic spins; as will be discussed later, the values of $\tau_0$ for superspin glasses tend to be several orders of magnitude larger. The observed scaling exponents are $\alpha \sim 0.1$ and $\mu \sim 0.9$; these values will be used in this paper for the comparison of superspin glass behavior and atomic spin glass behavior.

## 3. Results
### 3 (a) Evidence of a superspin glass transition
### (i) Susceptibility versus temperature measurements

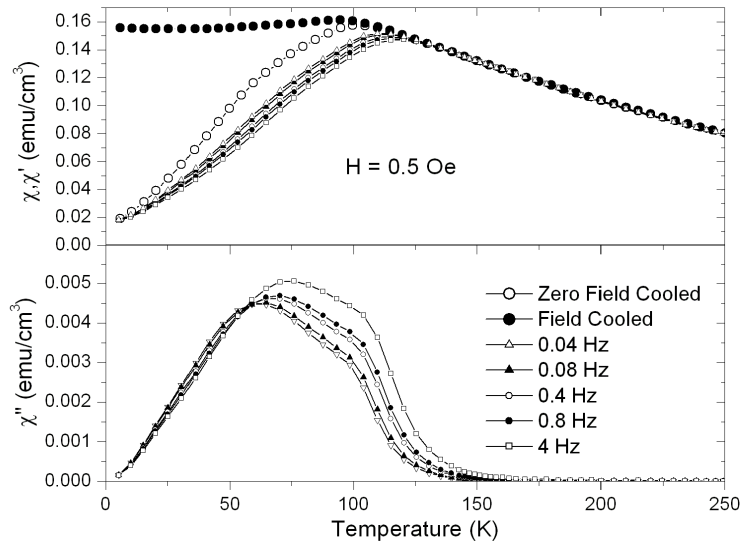

**Figure 2:** FC, ZFC and ac susceptibility vs. temperature of $\gamma$-Fe$_2$O$_3$, 35% in H$_2$O. Top: FC and ZFC susceptibility and the $\chi'$ component of the ac susceptibility measured at various frequencies. Bottom: The $\chi''$ component of the ac susceptibility measured at various frequencies.

Figure 2 shows the temperature dependence of the field cooled (FC) and zero field cooled (ZFC) dc susceptibilities as well as that of the in-phase ($\chi'$) and out-of-phase ($\chi''$) ac susceptibility. The dc FC and ZFC curves show typical Curie-Weiss behavior at high temperature. Below $T_g$ (=100 K) however, the FC susceptibility remains almost constant with temperature; close examination of the curve reveals a slight decrease below $T_g$ which is characteristic of a spin glass [11] and has also been observed for superspin glasses [5]. The ZFC susceptibility exhibits a pronounced peak at $T_g$ before decreasing with decreasing temperature, a feature also consistent with spin glass behavior [11]. We emphasize that the behavior observed here in this concentrated sample is significantly different from that observed in a more dilute ($\Phi = 10^{-2}$ %) system of the same particles [33]. In the dilute sample a peak is still observed in the ZFC susceptibility but at a lower temperature ($T_g$ = 67 K), and the FC susceptibility increases with decreasing temperature below $T_g$, a feature which is characteristic of the progressive freezing of nearly non-interacting superparamagnetic particles.

Assuming that interparticle interaction is negligible in the dilute sample, we can estimate the barrier energy, $E_B$, from $\tau = \tau_0 \exp^{E_B/k_B T}$ where $\tau_0$ is $10^{-9}$ s. This gives $E_B$ = 1695 $k_B$ = 2.3 × $10^{-20}$ J. This determination of the anisotropy energy gives a value slightly higher than that obtained by direct low temperature measurements of the anisotropy field on similar nanoparticles by FerroMagnetic Resonance [34]. However, it should be taken into account that the present sample polydispersity may increase $T_B$ significantly. The interparticle interaction energy in the concentrated ($\Phi$ = 35%) sample can be estimated from $\frac{E_{int}}{k_B} = T_{int} \approx \frac{\mu_0}{4\pi k_B} \mu m_s \phi$ [35]. We find $E_{int} \approx$ 1.11 × $10^{-21}$ J = 80 K. This is very close to the observed freezing temperature in the concentrated sample. Previous experiments on interacting nanoparticles have shown that freezing due to interactions occurs at temperatures equal to the interaction energy multiplied by a factor of the order of 1 to 2.5 [35], in good agreement with this result.

To further illustrate the transition towards a superspin glass state, the ac susceptibility versus temperature measured at 5 different frequencies from 0.04 to 4 Hz is shown in Figure 2. The $\chi'$ curves resemble the dc ZFC curve with a peak at approximately 100 K which shifts to higher temperatures with increasing frequency. This shift in peak temperature can be analyzed in terms of the Arrhenius law for a system of non-interacting magnetic particles, $\tau = \tau_0 \exp(E_a/k_B T)$, where $E_a$ is the anisotropy energy, $\tau$ is the inverse of the measurement frequency and $\tau_0$ is an attempt time. However by plotting $1/T_{peak}$ versus $\log 1/f$ (not shown) we find $\tau_0 = 10^{-19}$ s which is unphysically small. This result indicates a breakdown of the Arrhenius law: the relevant activation energy scale is here temperature dependent, which is expected for a system of strongly interacting particles encountering a spin glass–like transition [5] [36].

In order to analyze these data and test the hypothesis of the existence of a superspin glass transition, we have scaled our data using a critical law $\omega^{-1} = \tau^*(T_g(\omega)/T_g - 1)^{-z\nu}$ where $z$ is the dynamical critical exponent, $\nu$ the critical exponent associated with the correlation length and $\tau^*$ an attempt time which depends on the ratio $E_a/k_B T$, but whose temperature variation will be neglected in the narrow temperature range of our analysis. To constrain the analysis, we fixed $T_g$ to the value of the temperature at which the ZFC curve exhibits a pronounced maximum and optimized $\tau^*$ and $z\nu$. We find $\tau^* = 1 \times 10^{-9}$ s and $z\nu = 10$; the results are shown in Figure 3 in the form of a scaling plot. This value of $z\nu$ is consistent with that expected for an atomic spin glass [11, 15] and therefore strongly supports the existence of a phase transition towards a superspin glass state in this concentrated sample. Interestingly, the large value of $z\nu$ found here is close to values found for Ising spin glasses and larger than those found for Heisenberg spin glasses [37].

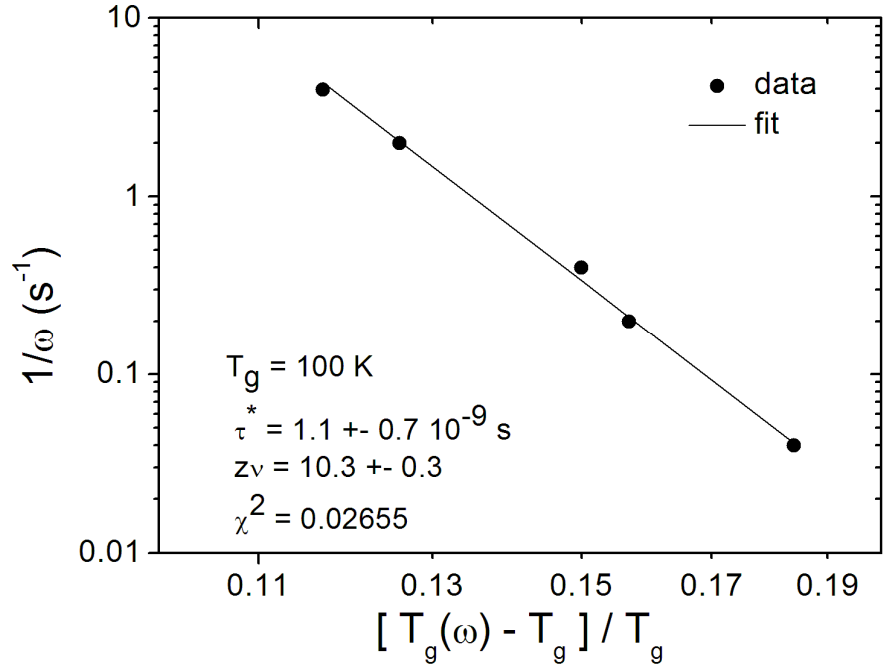

**Figure 3:** $1/\omega$ versus $(T_g(\omega)-T_g)/T_g$ on a log-log scale for a concentrated ($\Phi = 35\%$) dispersion of $\gamma$-$Fe_2O_3$ nanoparticles in water.

**(ii) Effect of the applied field: magnetic non linearities**

Figure 4 shows the dc susceptibility versus temperature measured in a range of dc applied fields varying from 0.3 to 60 Oe. Increasing the applied field above 5 Oe leads to a shift to lower temperature of the ZFC peak, accompanied by a decrease in magnitude of the susceptibility. The same effect is also observed in ac susceptibility curves measured in different applied dc fields (not shown). This behavior is known for atomic spin glasses [38-40] where, at temperatures above $T_g$, the magnetization is a function of the field: $M = (\chi_0 H) - a_3(\chi_0 H)^3 + a_5(\chi_0 H)^5 - ...$, which gives $\chi = \chi_0 - a_3\chi_0^3 H^2 + a_5\chi_0^5 H^4 - ... = \chi_0 + \chi_{nl}$ (where $\chi_{nl}$ denotes the non-linear susceptibility). For a (super)paramagnet the values of $a_i$ are independent of temperature whereas for a spin glass, critical behavior is observed with a power-law divergence at $T_g$ [41]. This divergence, hidden in the enhancement of $\chi_{nl}$ observed in this temperature region (Figure 4), can only be evidenced through a careful analysis of the temperature and field dependence of $\chi_{nl}$ which is beyond the scope of the present paper. A similar observation is made in [12] by Sahoo *et al.* who report a detailed investigation of the non-linear susceptibility in discontinuous $Co_{80}Fe_{20}/Al_2O_3$ multilayers which supports the existence of low-temperature spin-glass ordering. They also observed a decrease in the ZFC peak position with increasing applied magnetic field which, at low field, was found to give rise to an Almeida-Thouless line, further evidencing a spin-glass phase.

In our superspin glass sample, the deviation from a linear response occurs at much lower fields than in an atomic spin glass. This discrepancy can be explained by the fact that a typical superspin comprises ~ $10^4$ spins and therefore the Zeeman coupling will be much enhanced compared to that of an atomic spin glass.

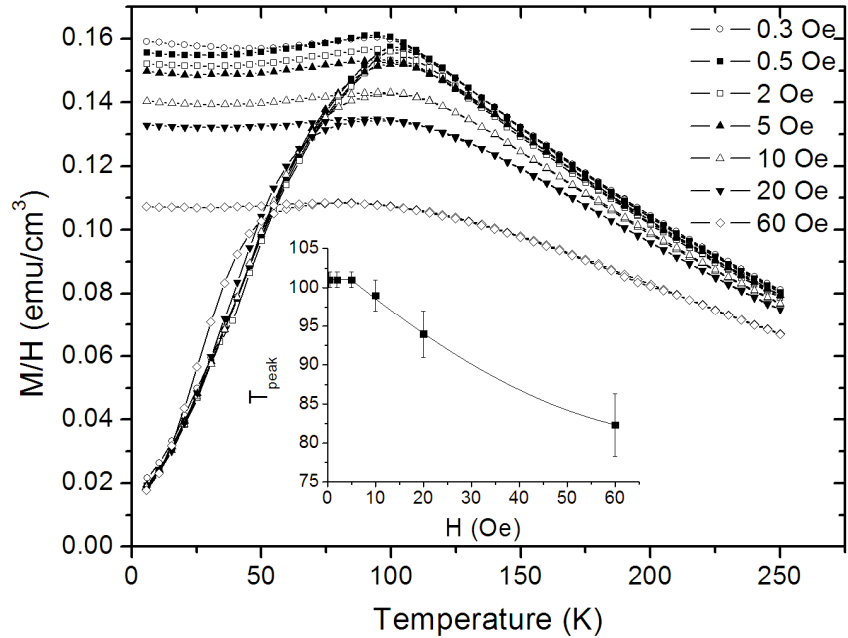

**Figure 4: M/H** versus temperature plot of a $\Phi = 35\ \%$ dispersion of $\gamma$-$Fe_2O_3$ nanoparticles measured with applied fields varying from 0.3 to 60 Oe. Inset: Variation of $T_g$ with applied magnetic field.

## 3 (b) Equilibrium/Out-of-equilibrium dynamics and scaling behavior
### (i) TRM experiments

We now focus on the slow dynamics in the superspin glass phase and present the results of the magnetization and ac susceptibility relaxation experiments performed at a measuring temperature $T_m = 0.7\ T_g$ (70 K) and an excitation field H = 0.5 Oe. Previous investigations of atomic spin glasses [28] have shown that *H* should remain small enough that the response of the sample remains in the linear regime to avoid an influence of the field on aging. This requirement is fulfilled in the present case (see fig.4).

Figure 5 shows the relaxation of the TRM, normalized to the field cooled magnetization ($M_{FC}$) for values of $t_w$ varying from around 1000 to 30 000 s. It can be seen that the relaxation depends on the value of $t_w$ as is observed for atomic spin glasses: this illustrates the aging character of the dynamics, the longer $t_w$, the slower the relaxation, indicating a 'stiffening' of the sample response during the waiting time.

It should be noted however that the spacing between the curves is not as great as is usually observed for an atomic spin glass, neither is the inflection point in the curves as clearly pronounced. The inflection point in the curves can be found by differentiating $M/M_{FC}$ with respect to log $t$ (as shown in figure 5 (b)) and, as for an atomic spin glass, we find that log $t_{infl}$ ~ log $t_w$ (see figure 5 (b), insert) indicating that $\mu$ ~ 1. However, when the relaxation curves are plotted against $t/t_w$ we do not observe an approximate scaling of the curves as found for atomic spin glasses. In order to achieve even a very rough scaling (as shown in the inset of Figure 5a) we plotted the TRM curves versus the scaling variable $\lambda/t_w^\mu$ (see section 2) with $\mu$ = 0.4 which is much smaller than is normally found for atomic spin glasses (0.7-0.9) and in contradiction with the $\mu$ ~1 behavior of the inflection points (Fig.5.b). The "straightness" of these curves suggests that there is a significant time-logarithmic contribution to the relaxation curves arising from a possible "superparamagnetic-like" relaxation of some of the particles in the sample, which is to some extent masking the spin glass-like behavior of the sample as a whole. The origin of this superparamagnetic behavior will be addressed in the section 4.

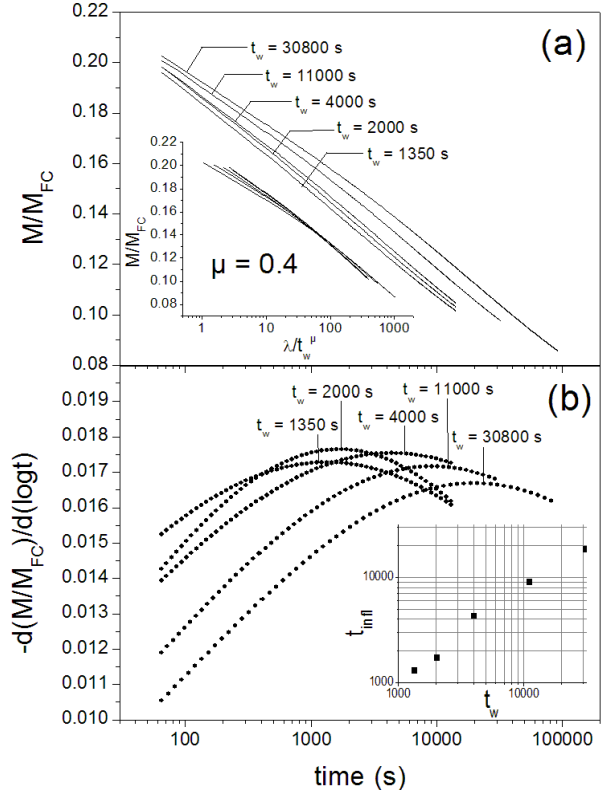

**Figure 5:** (a) TRM curves recorded at 0.7 $T_g$ normalized to the value of the FC magnetization for a variety of waiting times, $t_w$ ($H$ = 0.5 Oe). The inset is a scaling plot of the same curves as a function of the scaling variable $\lambda/t_w^\mu$, which is clearly inadequate (see text); (b) relaxation rate vs. time of the TRM curves shown in (a). The inset is a log-log plot of the inflection point time $t_{infl}$ vs. $t_w$.

In order to separate the superparamagnetic behavior of the sample from the spin glass behavior we have subtracted a term $-B\ln t/\tau_0$ from the TRM curves. Following this subtraction (see Figure 6 (a)) the $t_w$ dependence of the TRM curves now resembles that found for atomic spin glasses. By additionally subtracting an equilibrium part $(A(t/\tau_0)^{-\alpha})$ as is usual in the case of atomic spin glasses and plotting the resulting curves as a function of the scaling variable $\lambda/t_w^\mu$, it is possible to obtain a good scaling. This is shown in Figure 6 (b); the value of the scaling exponent $\mu$ of 0.90 and the scaling parameters ($A$ = 0.52 and $\alpha$ = 0.085) are within the range of those expected for an atomic spin glass. Note that the parameter values are additionally constrained by requirement of consistency with ac susceptibility measurements that will be discussed later.

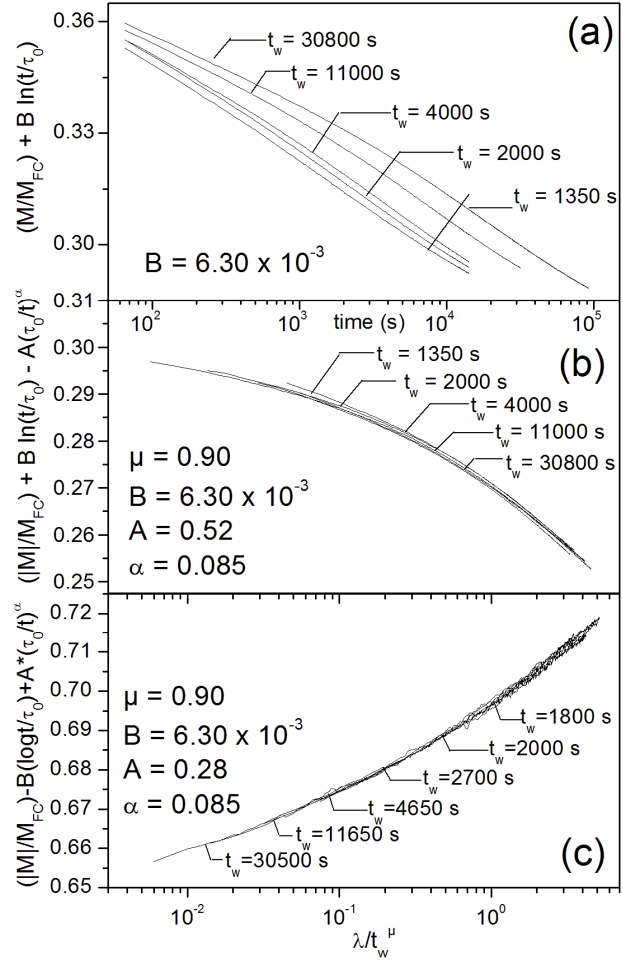

**Figure 6:** (a) TRM curves recorded at 0.7 $T_g$ for a variety of different waiting times, $t_w$ ($H$ = 0.5 Oe) following the subtraction of a $-B\ln(t/\tau_0)$ term; (b) Scaling of the TRM curves following the subtraction of a $-B\ln(t/\tau_0)$ term and an equilibrium part $A(t/\tau_0)^{-\alpha}$; (c) scaling of ZFCM curves following addition of a $-B\ln(t/\tau_0)$ term and an equilibrium part $A(t/\tau_0)^{-\alpha}$. See text for detail regarding the scaling procedures.

As previously suggested in [20], aging effects may in principle arise from a TRM protocol even in simple, non-interacting nanoparticle systems as during the field-cooling procedure most of the nanoparticles are frozen in an out-of-equilibrium state. In such a case, the initial field-cooled state evolves during the waiting time $t_w$, yielding an artificial $t_w$ dependence of the response functions. In order to confirm that the behavior we observe in our system is due to spin glass-like interactions between the particles and not to the trivial non-equilibrium state of the individual particles we have also performed zero field cooled magnetization (ZFCM) aging experiments as suggested in [20]. The advantage of using this protocol is that the sample is in zero field during cooling and $t_w$ and therefore is in an effective equilibrium in the absence of interactions between the particles. We have performed these ZFC relaxation measurements at the same measuring temperature as for the TRM and we find that scaling of the curves can be achieved with the same values of $\mu$, $B$ and $\alpha$. The parameter $A$ is reduced to from 0.52 to 0.28 (see Fig. 6 (c)). The data for the ZFCM relaxation curves are slightly noisier than in the case of the TRM; this arises from the fact that the ZFCM curves are recorded in the presence of a small applied field. Note that the aging phenomenon reported in the TRM case is still clearly present in the ZFCM protocol (with the same value of the scaling exponent $\mu$), which indicates that the aging phenomena reported come from the superspin glass phase.

For the sake of completeness, we should point out that, even though the scalings reported in Fig. 6(b) and (c) correspond to the best fits, another solution may be acceptable in which $\mu$, $A$ and $\alpha$ are those of the TRM scaling but B is decreased to $4.5 \times 10^{-3}$ - significantly smaller than in the TRM case ($6.3 \times 10^{-3}$). We propose to discard this solution, because more particles are ready to relax if a field is applied in a zero-field cooled state than in the mirror case of a field-cooled initial state (in which, due to the finite cooling rate, some of the particles are still frozen in a zero magnetization state and will not relax when the field is set to zero). Therefore $B_{ZTRM} < B_{TRM}$ is unlikely. In section 4(a) we shall discuss the effect of the two possible values of A on the consistency between *TRM*, *ZFCM* and ac experiments.

**(ii) ac susceptibility experiments**

To complement the previously described dc investigation of the out-of-equilibrium dynamics of our superspin glass sample we have also studied its aging behavior by ac susceptibility relaxation experiments. The measurement was carried out at $T_m = 0.7 \ T_g$ (70 K), in frequencies ranging from 0.04 to 8 Hz. As described in section 2(b), the sample was first cooled from above $T_g$ to the measuring temperature $T_m$ and then the susceptibility was measured over a time, $t_w$. The ac field used here was 0.5 Oe in all measurements to be consistent with the TRM experiments reported above.

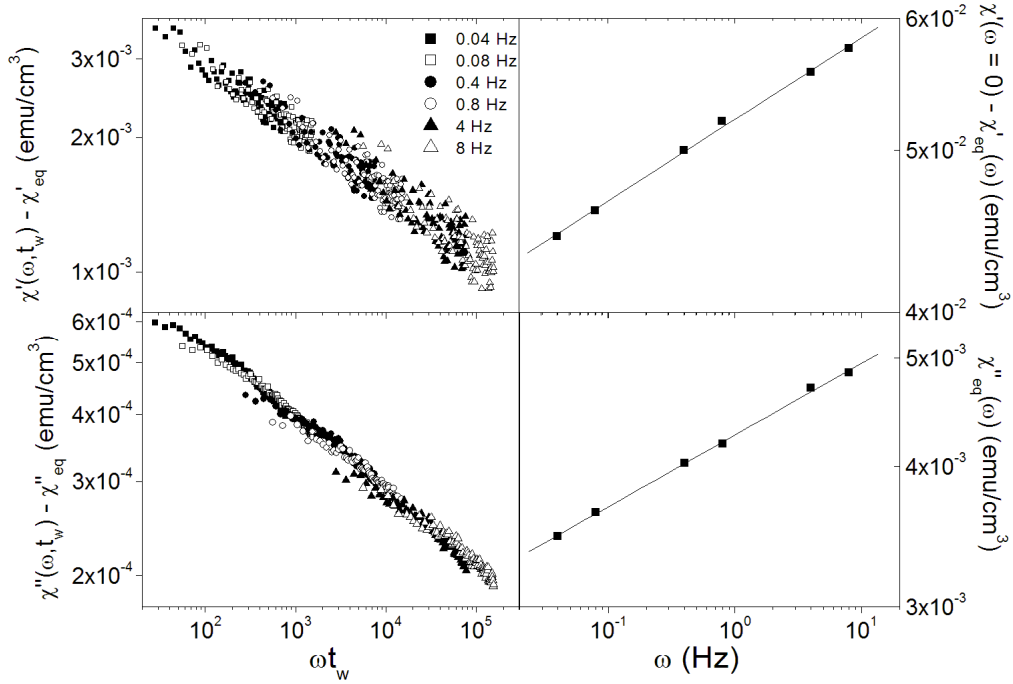

**Figure 7:** Left: Scaling plot of the aging parts of the χ' (top) and χ" (bottom) relaxations measured following a quench from $T_0$ (250 K) to $T_m$ (70 K). Right: Power law fit of the equilibrium part χ'$_{eq}$ (top) and χ"$_{eq}$ vs. frequency ω (see text).

As explained in section 2, the relaxation of the ac susceptibility of an atomic spin glass can be separated into two components, as for the TRM; an equilibrium stationary part, $\chi_{eq}(\omega)$ and an aging part, $\chi_{ag}(\omega, t_w)$, which behaves as a power law $\chi_{ag}(\omega, t_w) = A(\omega t_w)^{-b}$. In order to test this expression and the subsequent $\omega t_w$ scaling of the aging part of the ac susceptibility (equivalent scaling to the approximate $t/t_w$ scaling of the TRM) we have fitted $\chi(\omega, t_w)$ to $\chi_{eq}(\omega) + \chi_{ag}(\omega, t_w)$ using $\chi_{ag}(\omega, t_w) = A(\omega t_w)^{-b}$ and $\chi_{eq}(\omega)$ which depends only on $\omega$. The results for both $\chi_{ag}(\omega, t_w)$ and $\chi_{eq}(\omega)$ and for the in-phase and out of phase components are plotted in Figure 7 as a function of $\omega t_w$ and $\omega$ for the aging parts and the equilibrium parts respectively (left and right parts of Figure 7). As can be seen in this scaling plot, all the relaxation curves collapse onto a master curve $A(\omega t_w)^{-b}$ with b ~ 0.14, close to the value 0.2 found in atomic spin glasses. The equilibrium parts were fitted to expressions used in atomic spin glasses studies; $\chi'_{eq}(\omega) = \chi'(\omega=0) - A'_{eq}.\omega^{\alpha}$ and $\chi''_{eq}(\omega) = A''_{eq}.\omega^{\alpha}$, where $\alpha$ is the same as in the stationary part of the TRM. Figure 7 (right part) shows the power law fits of the corresponding equilibrium parts, $\chi'_{eq}(\omega)$ (top) and $\chi''_{eq}(\omega)$ (bottom). We emphasize here that these fits yield values of $\alpha$ significantly lower than those deduced from the TRM scaling. As we will show in the following discussion (Section 4), this discrepancy can be attributed to the superparamagnetic contribution revealed in the TRM measurements.

## 4. Discussion

### 4 (a) Consistency between dc and ac results

In order to obtain a consistent description of the scaling behavior of the response functions (both dc and ac) in our superspin glass sample and to achieve a coherent comparison with the scaling behavior of atomic spin glasses, we now present a more elaborate analysis of the TRM and ac susceptibility results taking into account the superparamagnetic contribution revealed in the TRM experiments. At the end of the analysis, we briefly consider the same question of consistency with the parameters given by ZFCM scaling. Finally, we discuss possible origins of the superparamagnetic contribution which highlight the intrinsic differences between superspin glasses and atomic spin glasses.

From the scaling procedure used in Figure 6(b) which takes into account a superparamagnetic contribution to the magnetization relaxation, the thermo-remanent magnetization recorded in TRM experiments can alternatively be written as, (with $\Gamma$ a numerical coefficient defined below):

$$\frac{M}{M_{FC}} = p\left[f\left(\frac{t}{t_w^\mu}\right) + A'\left(\frac{\tau_0}{t}\right)^\alpha\right] + (1-p)\left[1 - \Gamma \ln\left(t/\tau_0\right)\right] \quad (4)$$

where $p$ denotes the fraction of the system which behaves as a (super)spin glass (with "$f$" the aging function), $(1-p)$ is the fraction of the system which behaves as a superparamagnet. Comparing equation (4) with the scaling used for the TRM curves (see Figure 6(b)), one sees that $pA' \equiv A_{TRM}$, and that $(1-p)\Gamma \equiv B_{TRM}$. Performing the Fourier transform of equation (4) requires some precautions with respect to the superparamagnetic term. This term is simple to Fourier transform only when it behaves as a power law, namely as $(t/\tau_0)^{-\Gamma}$: this is true only when $t/\tau_0 \ll \exp(0.3/\Gamma)$. With the $\Gamma$ values involved below, this condition is not met over whole the experimental scale which extends up to $t/\tau_0 = 10^{10}$. To overcome this difficulty, we first note that the frequency interval [$f_{min}$ = 0.04 Hz, $f_{max}$ = 8 Hz] of the ac experiments is quite narrow (2.5 orders of magnitude) and we rewrite the superparamagnetic part of eq. (4) as $(1-p)[1-\Gamma \ln(t f_{max})]$ + $(1-p) \Gamma \ln(\tau_0 f_{max})$ which, over the time interval [$1/f_{max}$; $1/f_{min}$] accurately behaves as $(1-p)(t/\tau_0)^{-\Gamma}$ up to the constant term $(1-p) \Gamma \ln(\tau_0 f_{max})$. This allows us to get an explicit expression of the Fourier transform, whose validity is restricted to [$f_{min}$; $f_{max}$] but whose precision is ensured for the $\Gamma$ values involved below. We obtain, for the in phase component (with $G(x)$ the gamma function of $x$):

$$\frac{\chi' - \chi'_{aging}}{\chi'(\omega = 0)} = 1 - A_{TRM} G(1-\alpha) \cos\left(\frac{\pi\alpha}{2}\right) \times (\omega\tau_0)^\alpha + (1-p)G(1-\Gamma)\cos\left(\frac{\pi\Gamma}{2}\right) \times \left(\frac{\omega}{f_{max}}\right)^\Gamma \quad (5)$$

and for the out of phase component:

$$\frac{\chi'' - \chi''_{aging}}{\chi'(\omega = 0)} = A_{TRM} G(1-\alpha) \sin\left(\frac{\pi\alpha}{2}\right) \times (\omega\tau_0)^\alpha + (1-p)G(1-\Gamma)\sin\left(\frac{\pi\Gamma}{2}\right) \times \left(\frac{\omega}{f_{max}}\right)^\Gamma \quad (6)$$

In the classic case, as discussed in section 2, for atomic spin glasses, $p = 1$. As a result, $\chi''$ as well as $\chi'$ contain only one power law term whose exponent is that of the equilibrium term of TRM experiments. In our experiments, as the frequency range is quite narrow (2.5 orders of magnitude) and the exponents $\alpha$ and $\Gamma$ are both small, one can fit the measured $\chi'$ and $\chi''$ by simple power laws with effective exponents $\alpha'_{eff}$ and $\alpha''_{eff}$:

$$\frac{\chi' - \chi'_{aging}}{\chi'(\omega = 0)} = 1 - A'_{eff} (\omega\tau_0)^{\alpha'_{eff}} \quad (7)$$

$$\frac{\chi'' - \chi''_{aging}}{\chi'(\omega = 0)} = A''_{eff} (\omega\tau_0)^{\alpha''_{eff}} \quad (8)$$

From the measurements reported here we obtain $\alpha'_{eff}$ = 0.050 and $\alpha''_{eff}$ =0.064 with $A'_{eff}$ = 0.33 and $A''_{eff}$ = 0.026; the corresponding behavior must be compared with that expected from equations (5) and (6). Considering the uncertainties existing on some of the parameters (e.g. $A$, $\alpha$ (see below)) as well as the fact that the value of $p$ is not known *a priori*, we can check our results only through self-consistency, with the help of the following constraints:

(i) The scaling of the aging part of TRM experiments (see Figure 6(b)) can only be achieved when $B_{TRM} = 6.3 \times 10^{-3}$. Varying $B$ by as little as 15 % strongly degrades the scaling, whatever the values of other parameters ($A$, $\alpha$, $\mu$). We thus consider $B_{TRM}$ as fixed to the value given above and, since $p$ can be fairly well constrained *a priori* (see (iii) below), $\Gamma$ is not a free parameter; it is fixed by $\Gamma = B_{TRM}/(1-p)$.

(ii) In TRM experiments the aging part of the magnetization, denoted by the function $f$, must be positive. Therefore, from equation (4) and taking the scaled $M/M_{FC}$ value at the maximum $t$ measured (see Figure 6 (b)) we find that $(1-p) \leq 0.25$ which gives $p \geq 0.75$.

(iii) From the TRM scaling and from ac measurements we find $\alpha''_{eff} < \alpha_{TRM}$. Therefore, as equation (8) is effectively the sum of the 2 terms in equation (6) the exponent $\Gamma$ must be smaller then the exponent $\alpha''_{eff}$. This gives $\Gamma = B_{TRM}/(1-p) < \alpha''_{eff}$. Using $B_{TRM} = 6.3 \times 10^{-3}$ (as explained in (i)) we find $p \leq 0.90$. Therefore, $p$ must lie in the interval [0.75; 0.90].

(iv) The values of $A_{TRM}$ and $\alpha$ are only constrained by the requirement of a good scaling of the aging part of magnetization in TRM experiments (see Figure 6(b)). It is found that, once $\alpha$ is set, the uncertainty of $A$ around its optimal value is of only a few percent. Finally, the scaling in Figure 6(b) is only of good quality when $\alpha$ ranges between 0.085 (giving $A_{TRM}$= 0.52) and 0.15 (giving $A_{TRM}$ = 1.8).

Taking into account these constraints, the best agreement between ac experiments (equations (7) and (8)) and predictions from TRM experiments (equations (5) and (6)) is obtained with $\alpha = 0.085$ ($A_{TRM} = 0.52$), $p = 0.785$ and $B = 6.3 \times 10^{-3}$ (we have set $\tau_0 = 1$ ns). For this set of parameters, equation (6) (and respectively equation (5)) behaves effectively as a single power law of $\omega$ whose exponent approaches the $\alpha''_{eff}$ (respectively $\alpha'_{eff}$) value up to within 2 % (respectively 2 %). However, the absolute value predicted for the right hand side of equation (5) is 2% smaller than that directly obtained from AC measurements (eq. (7)). Lastly, the values calculated from eq. (6) are 25% smaller than those measured directly (eq. (8)) [42]. The fact that the relative difference between equations (6) and (8) is ten times larger than the corresponding difference between equations (5) and (7) comes from the fact that the out of phase response is typically 10 times smaller than the in phase response.

A similar analysis can be made by using the parameters of the ZTRM scaling of Fig. 6(c): $\alpha = 0.085$, $A_{ZFCM} = 0.28$, and $B = 6.3 \times 10^{-3}$. The best result is obtained with $p = 0.785$. In this case the agreement on exponents is still good (between 10% and 15%) but the fact that $A$ is smaller than in the TRM case degrades the agreement between the absolute values calculated from equations (5) and (6) and those directly measured (equations (7) and (8)). The relative difference in $1-\chi'$ is 15% while the values calculated from eq.(6) are twice smaller than those directly measured.

We emphasize that the frequency dependent part of $\chi'$ represents only 25 % of the total $\chi'$. As a consequence, even in the ZFCM case where the disagreement is the largest, the discrepancy of 15 % between equation (7) and equation (5) represents less than 4 % of the total measured ac signal. As far as we understand it, this 4 % difference might come (i) from the fact that the $\chi'(\omega = 0)$ value, entering in equations (7) and (8) is not directly measured but inferred with the *a priori* constraints that it must lie between the lowest $\omega$ ac measurement and the $\chi_{FC}$ value; (ii) from the fact that the coil used to produce the ac field is not the same as that used for TRM and ZFCM experiments and hence the absolute values of fields might slightly differ in these two cases. We finally conclude that the overall consistency between TRM/ZFCM measurements and ac experiments is satisfactorily checked for the above sets of parameters and that the extra superparamagnetic term $-B \ln(t/\tau_0)$ used in TRM experiments is of some importance also in ac experiments. We emphasize that this term is not visible when, instead of performing the full scaling of aging TRM, one focuses on $S(t) = dM(t)/d\ln(t)$ and looks for its maximum located at $t \approx t_w$. Therefore, the extra superparamagnetic term $-B \ln(\tau/\tau_0)$ might exist in other systems studied previously [10, 15, 24], since in these works the full scaling of aging TRMs was not carried out.

**4 (b) Physical origin of the superparamagnetic term**

We will now consider the possible physical origin of the extra superparamagnetic term $B \ln(t/\tau_0)$ used above. One possible idea is that, locally, the (dipolar) coupling constant $J_{ij}$ between neighboring nanoparticles might significantly fluctuate. Indeed in our case $J_{ij} \sim M_i M_j / r_{ij}^3$ where $M_i \propto V_i$ is the magnetic moment of the nanoparticle "$i$" of volume $V_i$ and $r_{ij}$ the distance between the two particles "$i$" and "$j$". In RKKY *atomic* spin-glasses, it was already noticed [43] that the high dilution of magnetic atoms, combined with the $J_{ij} \sim 1/r_{ij}^3$ behavior, results in a large distribution for $J$, decreasing at large $J$s as $P(J) \sim 1/J^{(1+\nu)}$ (with $\nu = 1$). The effect of such large distributions of $J_{ij}$ was studied in reference [43] and it was found that for $\nu < 2$ the physical behavior strongly differs from that of "canonical" spin glasses where $P(J)$ is a gaussian.

In our case, we have $J_{ij} \sim M_i M_j / r_{ij}^3$ but the fact that the volume fraction of nanoparticles is very high severely limits the fluctuations of $J$, since the fluctuations of $r_{ij}$ are much smaller than in the case of RKKY atomic spin glasses. However, the lognormal distribution of $d_i$, (the particle diameter, with a standard deviation for $\ln(d_i)$ given by $\sigma_d = 0.25$), results in a lognormal distribution of the volumes (and consequently the moments, $M_i$) whose standard deviation is $\sigma_M = 3*\sigma_d = 0.75$, yielding finally a lognormal distribution for $J$ with $\sigma_J = 0.69 \times 2^{1/2} = 1.06$, since $\ln(J)$ is the sum of the independent Gaussian variables $\ln(M_i)$ and $\ln(M_j)$. Thus $J$ is lognormally distributed, and not distributed as $P(J) \sim 1/J^{(1+\nu)}$ as in reference [43]. However the quite large value $\sigma_J = 1.06$ makes the distribution of $J$ very large and not that different from the case $P(J) \sim 1/J^{(1+\nu)}$ with $\nu < 2$. More precisely, this lognormal distribution of $J$ decreases with $J$ more slowly than $1/J^3$ for $J < J^*$ with $J^* \approx 17 J_{typ}$ where $J_{typ}$ is the most probable value of $J$. For $J > J^*$, $P(J)$ decreases more quickly than $1/J^3$, which means that our case does not fully correspond to the calculations of reference [43]. However, the fraction of couplings larger than $J^*$, given by the integral of $P(J)$ from $J^*$ to infinity, is less than 2.5%. This is why reference [43] should reasonably model our sample, up to a small approximation. Note finally that we have conservatively disregarded the (small) fluctuations of $r_{ij}$. However, a mere fluctuation of ±12 % of $r_{ij}$ around its mean value adds a further factor of 2 on the spreading of $J_{ij}$.

One important result of reference [43] is that for $v < 2$, the spin glass transition taking place at $T_g$ is peculiar: due to the large spreading of $J$s (especially in the region of high $J$ values) the nature of the spin-glass transition changes compared to the Gaussian case and becomes akin to a percolation transition, where only a fraction p of spins are involved in the spin glass state (with $p = (T_g/T–1)^\beta$ and $\beta = 0.5$ for 3D systems). The complementary fraction (1-$p$) is made of "fast spins", i.e. of spins not strongly enough coupled to their neighbors (small $J$s) to belong to the spin glass "backbone". *In atomic spin glasses*, these fast spins do not contribute to aging TRMs, but might be seen, on the contrary, as being responsible for the "quasi instantaneous" decrease of $M_{FC}$ when the field $H$ is cut. *In our case*, these fast spins might be what we have called the "superparamagnetic contribution" as even if it is not coupled enough to its neighbors to contribute to aging, a magnetic nanoparticle relaxes logarithmically slowly towards equilibrium, due to its anisotropy energy barrier. Finally, note that in our experiment we have $T/T_g = 0.7$ which yields a predicted value for $p = (T_g/T - 1)^\beta = 0.65$, in reasonable agreement to that used in the analysis of our experiments ($0.75 \leq p \leq 0.90$).

## 5 Conclusions

In this paper we have studied the low temperature superspin glass behavior of a concentrated frozen ferrofluid made of γ-$Fe_2O_3$ nanoparticles using SQUID magnetometry. We have focused on the out of equilibrium behavior of the superspin glass phase by studying the aging of both the TRM and ac susceptibility. It was found that the scaling laws normally applied to atomic spin glasses are also valid for our superspin glass sample and good agreement was found between the scaling parameters for the ac and dc relaxation curves. In order to achieve this scaling however, it was necessary to subtract a superparamagnetic contribution from the dc and ac response functions in the form of a term -$B \ln(t/\tau_0)$. We propose that this contribution, which was found consistently in the ac and dc measurements, arises from the large size distribution of the nanoparticles. This size distribution in turn results in a large distribution of coupling between neighboring particles enabling those nanoparticles which are only very weakly coupled to relax logarithmically. These results strongly support the existence of 'true' spin glass behavior in superspin glasses.